\documentstyle[mprocl]{article}

\def\be{\begin{equation}}
\def\ee{\end{equation}}
\def\bea{\begin{eqnarray}}
\def\eea{\end{eqnarray}}

\begin{document}

\title{HAWKING RADIATION AND HAWKING FLUX FROM SPHERICAL
REDUCTION}

\author{W.\ KUMMER}

\address{Institut f\"ur Theoretische Physik, Technische 
Universit\"at Wien,\\ Wiedner Hauptstrasse 8-10/136, 
A-1040 Vienna, Austria\\
E-mail: wkummer@tph.tuwien.ac.at}

\author{D.V.\ VASSILEVICH}

\address{Institut f\"ur Theoretische Physik, Universit\"at 
Leipzig, Augustusplatz 10\\
D-04100 Leipzig, Germany\\
E-mail: vassil@itp.uni-leipzig.de}  


\maketitle
\abstracts{ 
Using heat kernel techniques we show that the relation 
between Hawking temperature and radiation flux known from 
Einstein gravity in D dimensions can be reproduced from the 
spherically reduced action. A recent controversy regarding 
the $D=2$ anomaly for that case is discussed. The 
generalized effective Polyakov action in the presence of a 
dilaton field is presented.
}

\section{Introduction}

Hawking radiation in $D=4$ is regarded as a well-understood 
quantum theoretical feature of black holes (BH) or of other 
geometric backgrounds with an event horizon. On the other hand, 
the dilaton theory emerging after spherical reduction from 
Einstein gravity (SRG) and generalizations of those 
theories, only during the last years have been investigated 
in this connection. If the consideration of spherically 
reduced actions like 
\begin{eqnarray} 
\label{eq:ddim}
S_{\rm SRG} & = & \int \, d^2 x \; {\cal L}_{\rm SRG} \\
{\cal L}_{\rm SRG} & = & \sqrt{-g} \; e^{-2 \phi} \; 
\left\{ R + \frac{4 (D-3)}{(D-2)} \; 
(\nabla \phi)^2 - e^{\frac{4}{D-2}\, \phi} + 
\frac{1}{2} \, (\nabla f)^2 \, \right\}
\end{eqnarray}
obtained by the ansatz ($R$ and $\nabla$ refer to  
$g_{\mu\nu} (x)$ in 2D) 
\begin{equation}
(ds)^2 = g_{\mu\nu} dx^\mu dx^\nu - 
\frac{e^{\frac{4}{D-2}\, \phi}}{ (D-2) (D-3)} \; 
(d \Omega_{D-2} )^2
\end{equation}
from the D-dimensional Hilbert-Einstein action especially 
also in the quantum case \cite{KLV} should make any sense, it 
seems a necessary condition that this basic result is 
reproduced. The flux of radiation at infinity should be 
related correctly to the surface gravity at the horizon of 
the D-dimensional BH and the Hawking temperature 
\cite{Hawking75} 
following from the latter. 

In fact, already Christenson and Fulling \cite{CF77} had
shown by the simple use of energy momentum conservation
$\nabla_\mu T^{\mu\nu} =0$ in $D=2$ {\em minimal} coupling
of the scalars (i.e.\ with the factor $\exp\,(-2 \phi)$ in
front of $(\nabla f)^2$ removed) in (2) that, as expected,
the flux to $\cal J_+$ ($T_{uu}$ refers to the appropriate
light cone conformal coordinates, cf.\ (8) below)
\begin{equation}
\left. {T\,}_{uu}^{(min)} \right\vert_{J_+} = \frac{\pi}{12} 
\, T_H^2
\end{equation}
in terms of the Hawking temperature ($r_h$ is the radius of 
the horizon in asymptotic coordinates) 
\begin{equation}
T_H = \frac{D-3}{4\, \pi\, r_h} 
\end{equation}
precisely corresponds to the $D=2$ Stefan-Boltzmann law. The 
only necessary input to that computation had been the 2D 
conformal anomaly (one-loop quantum contribution of $f$ to 
the trace of $T^{\mu\nu}$ ) and the condition that $\left. 
T_{uu}\right|_{r_h} = 0$ which means finite flux in global 
(Kruskal) coordinates. However, in \cite{CF77} a complete 
derivation of this result for the S-wave part of $D=4$ 
(non-minimal coupling of $f$ in (1)) could not be achieved 
from the  $D=4$ anomaly, despite its close similarity to the case 
of minimal coupling in $D=2$. 

This problem lay dormant for many years until the authors of 
\cite{MuK94} noticed a surprisingly difficult situation: 
although their approach did not proceed through integrating 
the energy momentum conservation as in \cite{CF77}, but 
through a (Polyakov-type) effective action, as determined 
from the anomaly, they found that by taking that piece of 
$T_{uu}\vert_{{\cal J}^+}$  alone no Stefan-Boltzmann flux, 
but even a negative flux was created. This originated from 
the large negative contribution of ${T^{(\phi)}}^\mu_\mu $  are 
produced by the dilaton dependent part in the anomaly 
\begin{equation}
{T_\mu}^\mu = {{T\,}^{(min)}}_\mu^\mu + {T^{(\phi)}}_\mu^\mu
\end{equation}
for the nonminimally coupled scalars as in (1). They 
realized that another conformally noninvariant contribution 
(caused by the $\phi$-field dependence) must be added. By 
perturbative methods they argued that then in the complete 
result the negative contribution would be completely 
cancelled yielding the expected result (3). 

In 1997 this problem obtained some publicity because it was 
rediscovered by Bousso and Hawking \cite{BH}. Unfortunately 
in their calculation of the anomaly in $D=2$ these authors 
overlooked two essential points (namely, the dilaton dependence
of the diffeomorphism invariant measure and certain contributions
from the zero modes). Several contradicting results have been
reported in the literature\cite{BH,ChS,NoO,KLV1,Ich}. It was 
even claimed\cite{NoO} that 
there was an inherent ambiguity in one term of 
the conformal anomaly anomaly. This discussion was essentially
closed by the papers \cite{DOW,KLV2}. It became a common
practise to use the conformal anomaly derived in\cite{MuK94,KLV1}
(see e.g. \cite{Bousso:1998bn}).

When the diffeomorphism 
invariant scalar product of matter fields in $D$ dimensions 
with measure $\sqrt{-g_{(D)}}$ is reduced to $D=2$, from the 
factor of $(d\, \Omega_{D-2})^2$ in (2) the corresponding $D=2$ 
relation becomes
\begin{equation}
\langle f_1, f_2 \rangle = \int d^2\, x \; \sqrt{-g} \, 
e^{-2\, \phi} \; f_1\, f_2
\end{equation}
This dilaton dependence is nothing else but the 
necessary factor $1/r$ for S-waves. 

Before the work of \cite{BH} was published the present 
authors together with H.\ Liebl \cite{KLV1} had given the 
general result for the anomaly, allowing arbitrary 
nonminimal coupling of the scalars ($ \exp\, (-2\phi) \to 
\exp \, (-2 \, \varphi (\phi)$) in 
front of $(\nabla f)^2$ of (1) ) with arbitrary dilaton 
dependent measure ($ \exp (-\phi) \to \exp (-2\psi  
(\phi)$ )in (6). Employing the heat kernel technique the 
special case of SRG in \cite{MuK94} was confirmed and a 
well-defined anomaly had been obtained in contrast to 
\cite{BH} (cf.\ also \cite{KLV2}). 

\section{Stefan-Boltzmann flux for spherically reduced 
gravity}

Still the question had not found an answer, how to arrive at 
the correct flux for SRG without relying on perturbative 
methods as in \cite{MuK94}, when the present authors 
realized that an extended energy momentum argument, together 
with a novel application of the heat kernel technique for 
the missing piece in (6) allowed a complete (analytic) 
solution \cite{KV1}. 

In the presence of a dilaton field the argument of 
\cite{CF77} 
must be applied to a {\em different} energy momentum 
conservation law. The diffeomorphism invariance of the 
matter part (1) for the $f$-fields satisfying the e.o.m.\ 
$(\delta L^{(m)} / \delta f = 0)$ does not lead to 
$\nabla_\mu T^{\mu\nu} = 0$ but to 
 
\begin{equation}
\nabla_\mu\, {T^\mu}_\nu = - ( \partial_\nu \phi ) \; 
\frac{1}{\sqrt{-g}} \; \frac{\delta\, L^{(m)}}{\delta\, 
\phi}
\end{equation}
a result also noted in \cite{BF98}. Therefore, integrating 
(7) according to \cite{CF77} for the quantum effects ${\cal 
O}(\hbar)$ on both sides also required information on the 
``dilaton anomaly'' on the r.h.s.\ (with $L^{(m)}$ replaced 
by the one-loop effective action $W$ ) beside the usual 
anomaly contribution to the l.h.s. .

In conformal gauge ($u = t-z, v=t+z$) 
\begin{equation}
g_{\mu\nu} = e^{2\, \rho}\; du\, dv
\end{equation}
with $z$ related to the radial variable $r$ by $dr/dz = \exp 
2\,\rho $ we start from the well-known fact that the 
functional derivative of the active action $W$ yields the 
anomaly of (6). For the effective action $W$ (in Euclidean 
space) 
\begin{equation}
\exp \, W = \int \, \left( d \tilde{f} \, \sqrt[4]{g} 
\right) \; 
\exp \, \int d^2 x \; \sqrt{-g} \; \tilde f \; A \; \tilde f
\end{equation}
with a redefined $\tilde f = f\, e^{-\psi (\phi)} $ and where 
${\hat g}^{\mu\nu} $ is the resulting effective metric in 
\begin{equation}
\hat A = {\hat g}^{\mu\nu} \, \hat D_\mu \hat D_\nu - E\; ,
\end{equation}
the corresponding covariant derivative,  the anomaly is 
extracted in $\xi$-function regularization of the functional 
determinant from the (elliptic) operator $A$ by a {\em 
multiplicative} change of $\hat A$ (i.e.\ of ${\hat g\,}^{\mu\nu}$ 
by  $\delta k (x) $) 
\begin{eqnarray}
\delta W & = & - \frac{1}{2} \, \int \, d^2\, x \, 
\sqrt{g}\; \delta k (x) \; {T_\mu}^\mu \; = \nonumber \\
& = & - \frac{1}{48} \, {\rm Tr} \, \int d^2 x \, \sqrt{\hat g} 
\, \delta k (x) \, (\hat R + 6 \hat E)\; ,
\end{eqnarray}
where the second line is the standard result in the heat 
kernel technique \cite{GI}. In this way for a general theory (cf.\ 
the last paragraph of Section 1)
\begin{equation}
{T_\mu}^\mu = \frac{1}{24 \pi} ( R - 6 (\nabla \varphi)^2 + 4 
\Box \varphi + 2 \Box \psi )
\end{equation}
can be conputed, and in conformal gauge (9) with $\delta k = 
- 2 \delta \rho$
\begin{equation}
\frac{\delta W}{\delta \rho} = - \sqrt{-g} \; 
{T_\mu}^\mu\; ,
\end{equation}
where  $\sqrt{-g} R = - 2 \Delta \rho \quad ( \Delta = 
\eta^{\mu\nu}\, \partial_\mu \partial_\nu$) and 
$\sqrt{-g}\, \Box = \Delta$ may be used. 

The computation of the dilaton anomaly (r.h.s.\ of (7) with 
$L^{(m)} \to W$) starts from the identity in conformal gauge
\begin{equation}
\frac{\delta W \, (\rho, \phi)}{\delta \phi} \; = \; 
\int\limits_0^\rho \, d \rho' \; \frac{\delta^2 W (\rho', 
\phi)}{\delta \rho'\, \delta \phi} + 
\frac{W_0 (\phi)}{\delta \phi}\; ,
\end{equation}
where in the (functional) integral over $\rho$ in the first 
term, eq.\ (13) can be used. The last term with
\begin{equation}
\frac{\delta W_0}{\delta \phi} = \frac{\delta W_0}{\delta 
\varphi}\, \frac{d \varphi}{d \phi} + \frac{\delta W_0}{\delta 
\psi} \, \frac{\delta \psi}{\delta \phi}
\end{equation}
refers to flat space. It can be made amenable to the 
multiplicative variation needed for the heat kernel 
technique by replacing the determinant of the differential 
operator $A_0$ in the effective action by one half of the 
contribution from a ``fermionic'' operator related to 
(commuting) Majorana fields $\chi$ as (again in Euclidean space) 
\begin{equation}
W_0 = \frac{1}{4} \; \ln\; \int (d \chi) \; \exp 
( - \int d^2 x \; \chi\, D \tilde D \, \chi )\; ,
\end{equation}
because after partial integration 
\begin{equation}
\int\limits_x \chi\, D \tilde D \, \chi = \int\limits_x f \, A_0 \, f
\end{equation}
with $D = i \gamma^\mu \; e^\psi \partial_\mu\, 
e^{-\varphi}$ and 
$\tilde D = D ( \psi \leftrightarrow -\varphi)$. 
In this form in the trace of the $\xi$-regularization {\em 
multiplicative} changes by $\delta \varphi$ and $\delta 
\psi$ result from the corresponding variation, and the step 
analogous to the one from (11) to (12) is applicable (for 
details we refer to \cite{KV1}). 

In the resulting functional differential equations for $W$
\begin{eqnarray}
- 12 \pi \; \frac{\delta W}{\delta \varphi} & = & 
6 \eta^{\mu\nu} \, \partial_\nu ( \rho \, \partial_\mu\, 
\varphi ) + 2 \Delta \rho - 2 \Delta \psi - \Delta \varphi 
\; ,\nonumber \\
- 12 \pi \; \frac{\delta W}{\delta \varphi} & = & 
\Delta \rho - 2\, \Delta \varphi - \Delta \psi\; , \\
- 12 \pi \; \frac{\delta W}{\delta \rho} & = & 
- \Delta \rho - 3 \eta^{\mu\nu} \, 
(\partial_\mu \varphi) ( \partial_\nu \varphi) + 
2 \Delta \varphi + \Delta \psi\; , \nonumber
\end{eqnarray}
the variations $\delta \varphi$ and $\delta \psi$ are still 
independent. Integrating (8) in the conformal gauge 
(${T^\mu}_\mu =  4 e^{-2\rho}\, T_{uv} $, the only 
nonzero elements of the affine connections are 
$ {\Gamma_{vv}}^v 
 = 2 \partial_v\, \rho, {\Gamma_{uu}}^u = 2 \partial_u\, 
 \rho $, for 
a background $\rho = \rho (z)$ and stationary flux 
$\partial_u = - \partial_v = - \frac{1}{2} \partial_z $) one 
finds that the new piece for the physically most interesting 
application $\varphi = \psi = \phi$ of SRG {\em exactly 
cancels} the contribution from the $\phi$-dependent part of 
the conformal anomaly, apart from a total divergence, 
yielding 
\begin{equation}
T_{uu} = {T\,}_{uu}^{(min)} + \frac{1}{16\, \pi} \; 
\left[ \, 2 ( \partial_z \phi) ( \partial_z \rho) + 
2 \rho \, ( \partial_z \phi)^2 + ( \partial_z \phi)^2 - 
\partial^2_z \phi \; \right]^{r=r(z)}_{r=r_h}\; ,
\end{equation}
where, according to the Unruh vacuum condition $T_{uu}$ has 
been assumed to vanish at the horizon $r_h$. Eq.\ (20) holds 
for a general background $e^{2\rho (r)} = K (r)$, $dr/dz 
= K (r)$, with $L (r_h) = 0$ and dilaton field $\phi (r)$ 
and for all $r \geq r_h$. Thus for a D-dimensional BH with
\begin{equation}
K_{BH} = 1 - \left( \frac{r_h}{r} \right)^{D-3}
\end{equation}
and 
\begin{equation}
T_{uu} = {T\,}_{uu}^{(min)} + \frac{1}{16\, \pi} \; 
\frac{K^2}{r^2} \; \ln\, K/\mu
\end{equation}
at ${\cal J}_+ ( r \to \infty)$ only the (expected) result  
(4) remains. In (22) also the renormalization contribution 
(with factor $\ln\, \mu$) has been added \cite{GI} as another 
piece in the second term. In global coordinates with factor 
$K^{-2}$ the latter yields  a logarithmic divergence at the 
horizon which, nevertheless, still implies an integrable 
flux. 

Finally it should be emphasized that the functionally 
integrated effective action can be obtained by inspection 
from (19) (making it covariant by $\Delta \varphi \to 
\sqrt{-g}\, \Box$, $\Delta \rho \to  \sqrt{- g}\, R/2$ ) :
\begin{eqnarray}
W  & = & - \frac{1}{24\, \pi}\; 
\int d^2 x\, \sqrt{-g}\, \Big[ \,
- \frac{1}{4} \, R\, \Box^{-1}\, R + 
3 (\nabla \varphi)^2 \, \Box^{-1} \, R - 
R ( \psi + 2 \varphi) + \nonumber\\
&& \quad +  (\nabla \psi )^2 + (\nabla \varphi)^2 
+ 4 (\nabla^\mu\, \psi) ( \nabla_\mu 
\varphi ) \; \Big] \; + \; W^{(ren)}
\end{eqnarray}

For a generally nonminimally coupled scalar field $\varphi 
(\phi), \psi = \psi (\phi)$ it represents the (exact) 
generalization of the Polyakov action \cite{Pol} for $\psi = 
\varphi = 0$. However, we did not use this action in our 
argument, because it is derived from {\em local} (UV) 
quantum effects. To directly calculate a flux at ${\cal J}_+$ 
from (23), as in the approach of \cite{Bur99} 
 in our opinion introduces the need for further 
input for its asymptotic behavior (e.g.\ for 
$\Box^{-1}$). 

\section{Conclusion}

Our result passes the main test, the relation between 
Hawking temperature and flux at ${\cal J}_+$. The presence 
of the (mild) logarithmic singularity at horizon has been 
considered unphysical in \cite{BF2}. However, it seems 
closely related to the renormalization procedure in $D=2$, 
and it is also not forbidden by any $D=4$ calculation along 
the lines of our approach (which is sadly missing so far). 
Another problem is the transition to the spinor operator 
$D\tilde{D}$ in (18), a step which lacks complete 
mathematical rigor, although it gave in the end the 
completely acceptable central result. Thus further work will 
be needed, also in view of the ``dimensional 
reduction anomaly'', 
which in general may cause unphysical effects after 
spherical reduction \cite{Z}, although we believe them not 
to be obviously relevant in our example, because so far it 
has passed all consistency checks. 

\section*{Acknowledgements}

The authors are grateful for the support by Fonds zur 
F\"orderung der Wissen\-schaft\-lichen Forschung (Austrian 
Science Foundation), Project P12815-TPH, of the Alexander 
von Humboldt Foundation and of DFG 
project Bo 1112/11-1.

\eject


\vspace*{-9pt}

\section*{References}
\begin{thebibliography}{99}

\bibitem{KLV} W.\ Kummer, H.\ Liebl and D.\ V.\ Vassilevich, 
Nucl.\ Phys.\ B {\bf 493} 491 (1997); B {\bf 513} 723 (1998); 
B {\bf 544} 403 (1999); 
D.\ Grumiller, W.\ Kummer and D.V.\ Vassilevich, Nucl.\ Phys.\ B 
{\bf 580} 438 (2000); cf.\ also for a review the 
contribution of W.\ Kummer to Parallel Session AT1  
in this volume. 

\bibitem{Hawking75} S.\ W.\ Hawking, Commun.\ Math.\ Phys.\ {\bf 
43}, 199 (1975); W.\ G.\ Unruh, Phys.\ Rev.\ D {\bf 14}, 870 (1976). 

\bibitem{CF77} S.M.\ Christensen and S.A.\ Fulling, Phys.\ Rev.\ 
D {15} 2083 (1977). 

\bibitem{MuK94} V.\ Mukhanov, A.\ Wipf and A.\ Zelnikov, Phys.\ 
Lett.\ B {\bf 332}, 283 (1994). 

\bibitem{BH} R.\ Bousso and S.W.\ Hawking, Phys.\ Rev.\ D {\bf 
56} 7788 (1997); D {\bf 57} 2436 (1998). 

\bibitem{ChS}
T. Chiba and M. Siino, Mod. Phys. Lett. {\bf A 12} 709 (1997).

\bibitem{NoO}
S. Nojiri and S. Odintsov, Mod. Phys. Lett. {\bf A 12} 2083 (1997);
Phys. Rev. {\bf D 57} 2363 (1998).

\bibitem{KLV1} W.\ Kummer, H.\ Liebl and D.V.\ Vassilevich, 
Mod.\ Phys.\ Lett.\ A {\bf 12} 2683 (1997). 

\bibitem{Ich}
S. Ichinose, Phys. Rev. {\ bf D 57} 6224 (1998). 

\bibitem{DOW} J.\ S.\ Dowker, Class.\ Quant.\ Grav.\ {\bf 15} 1881 
(1998).

\bibitem{KLV2} W.\ Kummer, H.\ Liebl and D.V.\ Vassilevich, 
Phys.\ Rev.\ D {\bf 58} 108501 (1988). 

\bibitem{Bousso:1998bn}
R.~Bousso,
Phys.\ Rev.\  {\bf D58}, 083511 (1998).

\bibitem{KV1} W.\ Kummer and D.V.\ Vassilevich, Phys.\ Rev.\ D 
{\bf 60} 084021 (1999)\\
and Annalen der Physik (Leipzig) {\bf 8} 801 (1999). 

\bibitem{BF98} R.\ Balbinot and A.\ Fabbri, Phys.\ Rev.\ D {\bf 
59} 044031 (1999).

\bibitem{GI} P.B.\ Gilkey, J.\ Diff.\ Geom.\ {\bf 10} 601 (1975) 
and ``Invariance Theory, the Heat Equation, and the 
Atiyah-Singer Index Theorem'', CRC Press, Boca Raton 1994.

\bibitem{Pol} A.\ M.\ Polyakov, Phys.\ Lett.\ B {\bf 103} 207 
(1981). 

\bibitem{Bur99} M.\ Buric, A.\ Mikovic and V.\ Radovanovic, 
Phys.\ Rev.\ D {\bf 59} 084002 (1999).

\bibitem{BF2} R.\ Balbinot and A.\ Fabbri, Phys.\ Rev.\ D {\bf 
59} 044031 (1999); Phys.\ Lett B {\bf 459} 112 (1999). 

\bibitem{Z} A.\ Frolov, P.\ Sutton and A.\ Zelnikov, Phys.\ Rev.\ 
D {\bf 61} 02421 (2000); P.\ Sutton, Phys.\ Rev.\ D {\bf 62} 
044033 (2000). 

\end{thebibliography}
\end{document}